\documentclass[twocolumn,floats,floatfix,nofootinbib,aps,pra]{revtex4-2}

\usepackage{amsbsy,amssymb,amsmath}
\usepackage{physics}
\usepackage{color,calc}
\usepackage{bm}
\usepackage{array}
\usepackage{booktabs}
\usepackage{graphicx}
\usepackage{graphics}
\usepackage{eepic,epsfig}
\usepackage[table, dvipsnames]{xcolor}
\usepackage{colortbl}
\usepackage{subcaption}

\usepackage[breaklinks,hidelinks,colorlinks=true,linkcolor=blue,citecolor=blue,filecolor=blue,urlcolor=blue]{hyperref}

\usepackage[dvipsnames]{xcolor}
\newcommand{\comment}[1]{}

\def\be{ \begin{equation} }
\def\ee{ \end{equation} }
\def\bea{ \begin{eqnarray} }
\def\eea{ \end{eqnarray} }
\def\bse{ \begin{subequations} }
\def\ese{ \end{subequations} }
\def\ba{ \begin{array} }
\def\ea{ \end{array} }
\def\bt{ \begin{tabular} }
\def\et{ \end{tabular} }

\def\i{\,\text{i}}

\def\i{i}

\def\d{\text{d}}

\def\U{\mathbf{U}}

\def\i{{\rm{i}}}
\def\f{{\rm{f}}}

\long\def\/*#1*/{}

\begin{document}

\title{Ultrabroadband, ultranarrowband and ultrapassband composite polarisation half-wave plates, ultrabroadband composite polarisation pi-rotators and on the quantum-classical analogy}

\author{Hayk L. Gevorgyan\textsuperscript{\hyperref[1]{1},\hyperref[2]{2},\hyperref[3]{3}}}
\email{hayk.gevorgyan@aanl.am}
\affiliation{
\phantomsection\label{1}{\textsuperscript{1}Center for Quantum Technologies, Faculty of Physics, St. Kliment Ohridski University of Sofia, 5 James Bourchier Blvd., 1164 Sofia, Bulgaria}\\
\phantomsection\label{2}{\textsuperscript{2}Division for Quantum Technologies, Alikhanyan National Laboratory (Yerevan Physics Institute), 2 Alikhanian Brothers St., 0036 Yerevan, Armenia}\\
\phantomsection\label{3}{\textsuperscript{3}Experimental Physics Division, Alikhanyan National Laboratory (Yerevan Physics Institute), 2 Alikhanian Brothers St., 0036 Yerevan, Armenia}}

%TODO 1) wait for phase gate publication

\date{\today }

\begin{abstract}
Composite pulses, which produce ultrabroadband, ultranarrowband and ultrapassband (\(x-\), \(y-\)) rotations by \(\theta = \pi\) on the Bloch-Poincaré sphere, are presented. The first class plays a role for design of achromatic polarisation retarders, when the second class corresponds to chromatic polarisation filters. The third class is an assortment of the above two classes. 

Besides, composite pulses, which produce ultrabroadband \(z-\) rotations by \(\zeta = \pi\) on the same sphere, are presented. These phasal pulses coincide with achromatic polarisation \(\pi\) rotators. 

On the quantum-classical analogy, we obtain ultrarobust, ultrasensitive and ultrasquare quantum control of a X gate and ultrarobust quantum control of a Z gate.

\end{abstract}

\maketitle

%%%%%%%%%%%%%%%%%%%%%%%%%%%%%%%%%%%%%%%%%%%%%%%%%%%%%%%%%%%%%%%%%%%%%%%%%%%%%%%%%%%%%%%%%%%%%%%%%%%%%%%%%%%%%%%%%%%%%%%%%%%%%%%%%%%%%%%%%

%()%()%()%()%()%()%()%()%()%()%()%
%()%()%()%()%()%()%()%()%()%()%()%
%()%()%()%()%()%()%()%()%()%()%()%

\section{Introduction}\label{Sec:intro}

%()%()%()%()%()%()%()%()%()%()%()%
%()%()%()%()%()%()%()%()%()%()%()%
%()%()%()%()%()%()%()%()%()%()%()%

In comparison to the other quantum control methods, composite pulses (CPs) is efficient and versatile as follows from it's classification into broadband (BB), narrowband (NB) and passband (PB) classes \cite{wimperis1994}. In addition, subclassification of the BB class into constant and variable rotations \cite{levitt1986, levitt2007} opens new perspective for full-matrix optimization, i.e. gate sequences \cite{gevorgyan2021, gevorgyan2024}.   

All the artillery of ultrahigh-fidelity (flat-top), broadband, constant rotation (full SU(2) matrix optimized) CPs is shown in \cite{gevorgyan2021}, where the well-known BB1 from Wimperis is one of the representatives of this subclass. CPs of this subclass, in contrast to altering-top BB2, maintain 99.99\% fidelity (ultrahigh), viz. infidelity is below $10^{-4}$ error of quantum computation benchmark required in quantum computing (QC) and quantum information (QI). On the contrary, in polarisation optics (PO), ultrabroadness is more important than ultrahigh-fidelity. Nevertheless, CPs in \cite{gevorgyan2021} can be used also to design achromatic ultrahigh-fidelity constant rotational half-, quarter- and arbitrary-wave plates with an arbitrary phase retardation.

Based on the concept of CPs applied for rotations on the Bloch sphere, Ardavan proposed to use BB1 and BB2 sequences for polarisation retarders, i.e. rotations on the Poincaré sphere \cite{ardavan2007}. He already found that these stacked composite retarders \cite{pancharatnam1955, koester1959, title1975, mcintyre1968} in almost all cases outperform the conventional compound-type retarders \cite{clarke1967, beckers1971, beckers1972}. 

Existence of BB2 and NB2 sequences leads to the idea of altering CPs, which can improve the feature (BB or NB) of the pulse at the expense of precision benchmark due to alternations (inflection points) on the top (BB) or on the bottom (NB) of the precision measure, in our case --- the conversion efficiency in polarisation optics or, mathematically equivalent, the errant transition probability in quantum optics. We call these new subclasses of variable rotational CPs as ultrabroadband and ultranarrowband, respectively. Moreover, we examine also the subclass of ultrapassband CPs, which have alternations both on the top and on the bottom, improve both BB and NB features, i.e. PB feature. 

With the novel method (see Sec.~\ref{Sec:derivation-ultra}) we have derived ultrabroadband, ultranarrowband \cite{gevorgyan2022} and ultrapassband CPs, when $\theta=\pi$. These CPs are useful in the applications, where high-accuracy (about 90\%) is enough, although higher precision can be achieved increasing number of pulses, due to the abovementioned method. Besides PO, they can be used for high-fidelity ultrarobust or ultrasensitive control and population transfer, which we hope to be useful in quantum computing --- for ultrarobust (but not ultrahigh-precision) quantum operations and in quantum sensing --- for ultrasensitive (but not ultrahigh-precision) local addressing of trapped ions and atoms\footnote{Our paper is mainly a proposal for ultra-composite pulses which primarily have ultra-properties, i.e. ultrabroadness, ultranarrowness or ultrapassbandness, Moreover, our paper shows that ultrahigh-precision theoretically can be achieved by using very long composite pulses, for the theoretical foundation of which the very powerful supercomputer will be necessary. In our context, ultrahigh-precision means that alternations should be limited to a precision difference of $10^{-4}$.}. For example, with the same $T=5\pi$ run-time\footnote{Also called an operation time. In the case of rectangular laser pulses in quantum optics, which is commonly used in composite pulses technique, $\theta_i \propto T_i$, where $T$ is the pulse's operation time. Since the pulses in a sequence are applied instantaneously one after the another, the total operation time can be measured by $T = \sum\limits_{i = 1}^{N} {\theta_i}$, considering that all the pulses have the same Rabi frequency. In polarisation optics, the phase shift of each waveplate is linearly proportional to it's thickness $\varphi_i \propto L_i$, since we have they are made from the same material, i.e. $n_e$ and $n_o$ are the same for different waveplates in the sequence, and the same light is propagating through them. Time of operation of a waveplate is a time of propagation of the light's both ordinary and extraordinary components through it, i.e. proportional to the thickness of a waveplate $L_i \propto T_i$. So, the total run time of the stacked composite waveplate can be measured by $T = \sum\limits_{i=1}^{N} {\varphi_i}$.}, our five-pulse ultra-BB and ultra-NB CPs outperform the well-known BB2 and NB2 pulses, respectively. This was expected as the number of alternations is higher in the case of our CPs.

% Existence of BB2 and NB2 sequences leads to the idea of alternating CPs, which improve the feature (BB or NB) of the pulse at the expense of precision due to alternations on the top (BB) or on the bottom (NB) of the fidelity or the other precision measure, for example, errant transition probability. 

Originally, CPs are derived for $\theta$ rotations ($x-$, $y-$rotations or mixed rotations with arbitrary $\phi$), to which all the abovementioned is mainly dedicated. Here, besides the rotational $\theta = \pi$ CPs, we took also into account $z-$rotations on the Bloch-Poincaré sphere, which we call as phasal $\zeta$ CPs. We adapted our novel method for derivation of ultrabroadband phasal composite $\zeta = \pi$ pulses.

This paper is organized as follows. Sec.~\ref{Sec:rot} is dedicated to the rotation gates in the context of quantum optics and quantum computing. Sec.~\ref{Sec:quantum-classical} presents Jones matrices in the context of polarisation optics and quantum-classical analogy. In Sec.~\ref{Sec:CPs} we explain composite pulses approach in the context of quantum control and analogous polarization control. In Sec.~\ref{Sec:derivation-ultra} we disclose derivation methods for obtaining rotational $\theta = \pi$ (Subsubsec.~\ref{Subsubsec:rot-pulses}) and phasal $\zeta = \pi$ (Subsubsec.~\ref{Subsubsec:phas-pulses}) pulses. Secs.~\ref{Sec:ultrabb},~\ref{Sec:ultranb} and~\ref{Sec:ultrapb} deliver results on ultrabroadband, ultranarrowband and ultrapassband rotational $\theta = \pi$ composite pulses, when Sec.~\ref{Sec:ultrabbph} yields results for ultrabroadband phasal $\zeta = \pi$ composite pulses. Finally, Sec.~\ref{Sec:concl} presents the conclusions.

\section{Rotation gates}\label{Sec:rot}
Basically starting from TDSE (a time-dependent Shr{\"o}dinger equation) for a two-level system, one can reach the evolution operator for a single-qubit, which is called Rabi rotation gate from AMO (atomic, molecular and optical) devices in experimental QC \cite{chen2006, nielsen2000}, or theta pulse in nuclear magnetic resonance (NMR) \cite{wimperis1994}. Thus, each pulse in a CP sequence is considered resonant and hence it generates the propagator
\be\label{U1}
\U (\theta,\phi) = \left[ \begin{array}{cc} \cos(\frac{\theta}{2}) & -i e^{i\phi} \sin(\frac{\theta}{2}) \\ -i e^{-i\phi} \sin(\frac{\theta}{2}) & \cos(\frac{\theta}{2})  \end{array}  \right],
\ee
where $\theta=\int_{t_\i}^{t_\f}\Omega(t)\d t$ is the temporal pulse area (the rotation angle), $\Omega(t)$ is a Rabi frequency and $\phi$ is the phase of a coupling. SU(2) symmetry is a character for the physical-level gates \cite{chen2006, maslov2017} in experimental QC in contrast to U(2) quantum gates \cite{nielsen2000} in theoretical QC. However, from a physical point of view it is more natural to use SU(2) gates, which have det = 1.

\section{Jones matrices and on the quantum-classical analogy} \label{Sec:quantum-classical}
The Poincar\'e sphere shares much in common with the Bloch sphere: both operations --- the evolution matrix of quantum two-state system and Jones matrix for a retarder in $L R$ polarisation basis (with a phase shift $\varphi = \frac{2\pi}{\lambda} (n_e - n_o) L$, and rotated at an angle $\eta$) represent rotations on the phantom spheres, which are the geometrical description of the space of states (the first case --- quantum states, the second case --- polarisation states). Note that in both cases, besides the pure states (($\ket{0}$, $\ket{1}$) and ($\ket{L}$, $\ket{R}$)), we may have an infinite number of superposition states of them, hence, we have a Bloch-Poincar\'e to describe them and operations, i.e. rotations between them. \par
Jones polarisation matrix for a retarder with a phase shift $\varphi$ (the phase shift applied between the ordinary and the extraordinary ray passing through the retarder) and rotated at an angle $\eta$ (the rotation angle of the retarder’s optical axis) is given as (on the left-right circular polarisation basis)
\be\label{J}
\textbf{J}_{\eta}(\varphi) = \left[ \begin{array}{cc} \cos\left( \frac{\varphi}{2} \right) & i \sin\left( \frac{\varphi}{2} \right) e^{2 i \eta} \\ i \sin\left( \frac{\varphi}{2} \right) e^{-2 i \eta}  & \cos\left( \frac{\varphi}{2} \right) \end{array}\right] ,
\ee
note, that here, $\eta$ in PO differs from $\theta$ rotation parameter in NMR and QC.\par
For example, half- and quarter-wave plates rotated at an angle $\eta$, i.e. $\left(\lambda/2 \right)_{\eta}$, $\left(\lambda/4 \right)_{\eta}$, are described by $\textbf{J}_{\eta}(\pi)$ and $\textbf{J}_{\eta}(\pi/2)$ respectively.\\
The ideal half-, quarter- and arbitrary-wave plates are described with Jones polarisation matrix $\textbf{J}_{0}(\varphi)$ in the LR basis (up to a global phase factor):

\be\label{J0half}
\textbf{J}_{0}(\pi) = \left[ \begin{array}{cc} 0 & i  \\ i  & 0 \end{array}\right] 
\ee\par

\be\label{J0quarter}
\textbf{J}_{0}(\pi/2) = \tfrac{1}{\sqrt{2}} \left[ \begin{array}{cc} 1 & i  \\ i  & 1 \end{array}\right] 
\ee\par

\be\label{J0}
\textbf{J}_{0}(\varphi) = \left[ \begin{array}{cc} \cos\left(\frac{\varphi}{2} \right) & i \sin\left(\frac{\varphi}{2}\right) \\ i \sin\left(\frac{\varphi}{2}\right) & \cos\left(\frac{\varphi}{2}\right) \end{array}\right] 
\ee\par

A Jones matrix for a rotator in the LR basis represents
\be\label{Jt}
\textbf{J}_{\eta} = \left[ \begin{array}{cc} e^{i \eta} & 0 \\ 0 & e^{-i \eta} \end{array}\right] . 
\ee\par

Due to symmetry $i \rightarrow -i$, and transformations $\varphi \rightarrow \theta$, $\eta \rightarrow \phi/2$ from PO to NMR QC, we deal with the same mathematical framework
\begin{itemize}
\item
polarisation retarder is equivalent to x-rotation or quantum rotation gate, 
\item 
polarisation rotator is equivalent to z-rotation or quantum phase gate. 
\item 
Thus, one can apply the results from QC into the PO and vice versa, especially to use quantum control techniques and share knowledge between different areas. We see quantum-classical analogy of the rotations on the Bloch-Poincar\'e spheres. CPs (composite rotations) is interdisciplinary technique.
\item 
To adapt the results from NMR QC to PO, it is necessary to use $\theta_i \rightarrow \varphi_i$ and to change $\phi_i \rightarrow \pm 2\eta_i$ in expressions ~\eqref{U1} and ~\eqref{J} (sign is arbitrary as the composite phases with negative sign are also solutions in the symmetric sequences). So, the halfed composite phases are necessary to use for $\eta_i$-s in the PO.  
\end{itemize}\par
Henceforward, we will use NMR QC terminology and notation, and the results for PO can be obtained by abovementioned way.

\section{Composite pulses approach}\label{Sec:CPs}
Composite pulses technique requires a finite train of pulses with the well-defined composite phases (the phases of the couplings $\phi_{k}$ in QC or the rotation angles of the retarders' optical axes $\eta_{k}$ in PO) and temporal pulse areas $A_k$ in QC (or the phase shifts $\varphi_k$ applied between the ordinary and the extraordinary ray passing through the retarders and depending on the wavelength in PO) 
\begin{equation}\label{U^N}
\begin{split}
& \textbf{u} = \textbf{U}_{\phi_{N}}(\theta_N) \cdots \textbf{U}_{\phi_{3}}(\theta_3) \textbf{U}_{\phi_{2}}(\theta_2) \textbf{U}_{\phi_{1}}(\theta_1) \quad \rightarrow \\
& \quad \rightarrow \quad \textbf{J}_{\theta_{N}}(\varphi_N) \cdots \textbf{J}_{\theta_{3}}(\varphi_3) \textbf{J}_{\theta_{2}}(\varphi_2) \textbf{J}_{\theta_{1}}(\varphi_1).
\end{split}
\end{equation}

Considering pulse area error $\theta \rightarrow \theta(1+\epsilon)$ (equivalent to phase shift deviation from its actual value $\varphi \rightarrow \varphi(1+\epsilon)$), errant overall propagator is the SU(2) matrix, being chronological multiplication of type \eqref{U^N}
\begin{equation}\label{eq-u}
\begin{split}
& \quad \quad \quad \quad \textbf{u} (\epsilon) = \left[ \begin{array}{cc} u_{11}(\epsilon) & u_{12}(\epsilon) \\ -u_{12}^{\ast}(\epsilon) & u_{11}^{\ast}(\epsilon) \end{array}\right], \\ 
& p(\epsilon) = 1-|u_{11}(\epsilon)|^2 = |u_{12}(\epsilon)|^2 = \sin^2(\theta_{\epsilon}/2) \rightarrow \\
& \quad \quad \quad \quad \rightarrow I(\varphi_\epsilon) = \sin^2(\varphi_\epsilon/2),
\end{split}
\end{equation}
where $u_{11}(\epsilon)$ and $u_{12}(\epsilon)$ are the complex-valued Cayley-Klein parameters satisfying $|u_{11}(\epsilon)|^2+|u_{12}(\epsilon)|^2=1$. Errant transition probability $p(\epsilon)$ is 1 at the centre of bandwidth ($\epsilon = 0$) --- in QC the qubit state completely transfers from $\vert 0 \rangle$ to $\vert 1 \rangle$ due to $\pi$-rotation on a Bloch sphere (composite NOT-$\sigma_x$ gate). This is analogous to the conversion efficiency $I(\varphi_\epsilon)$ equal to 1 in PO --- the polarisation state completely converts from $\vert L \rangle$ to $\vert R \rangle$ due to $\pi$-rotation on the Poincar\'e sphere (composite half wave-plate). Note that a choice of the initial state is arbitrary. For example, the same rotation is applicable to $\ket{H}$ to $\ket{V}$ conversion (transition).

\section{Derivation Methods} 
\label{Sec:derivation-ultra}

So, errant overall propagator is SU(2) matrix
\be\label{eq-U-n}
\U_n (\epsilon) = \left[ \begin{array}{cc} \mathcal{U}_{11}(\epsilon) & \mathcal{U}_{12}(\epsilon) \\ -\mathcal{U}_{12}^{\ast}(\epsilon) & \mathcal{U}_{11}^{\ast}(\epsilon) \end{array}\right],
\ee
where $\mathcal{U}_{11}(\epsilon)$ and $\mathcal{U}_{12}(\epsilon)$ are the complex-valued Cayley-Klein parameters satisfying $|\mathcal{U}_{11}(\epsilon)|^2+|\mathcal{U}_{12}(\epsilon)|^2=1$.
We set their zero-error values to the target values,
\be\label{eq-theta}
\mathcal{U}_{11}(0) = \cos(\theta/2),\quad \mathcal{U}_{12}(0) = -i \sin(\theta/2) \exp(i\phi),
\ee
for rotational $\theta$ pulses, or
\be\label{eq-phi}
\mathcal{U}_{11}(0) = \exp(-i \zeta/2),\quad \mathcal{U}_{12}(0) = 0,
\ee
for phasal $\zeta$ pulses.\par

Taking Eqs.~\eqref{eq-theta} and \eqref{eq-phi} as a guide, let's consider the general form for general composite rotation
\be\label{eq-theta-err-main}
\U_n (\epsilon) = \left[ \begin{array}{cc} \exp(- i\zeta_{\epsilon}/2) \cos(\theta_{\epsilon}/2) & -i \sin(\theta_{\epsilon}/2) \exp(i\phi_{\epsilon}) \\ -i \sin(\theta_{\epsilon}/2) \exp(-i\phi_{\epsilon}) & \exp(i\zeta_{\epsilon}/2) \cos(\theta_{\epsilon}/2) \end{array}\right],
\ee
where $\theta_{\epsilon}$ is errant rotation angle and arranges $x-$ or $y-$rotations, i.e. rotational $\theta$ pulses or rotation gates, $\phi_{\epsilon}$ is errant relative phase angle and provides the turns from $x-$ to $y-$rotation, $\zeta_{\epsilon}$ is errant phase-shift angle (sometimes called geometric phase angle) and arranges $z-$rotations or phase-shift gate up to global phase in the case of phasal $\zeta$ pulses and also corresponds to Berry phase (originally examined in cyclic adiabatic processes) alternative in conventional CPs or rotations, i.e. rotational $\theta$ pulses. For rotational $\theta$ pulses parameters follows $\theta_{\epsilon=0} = \theta$, $\zeta_{\epsilon=0} = 0$ and $\phi_{\epsilon=0} = \phi$ ($\phi = 0$ is the case for ideal $\theta$ pulse), and for phasal $\zeta$ pulses parameters are equal $\zeta_{\epsilon=0} = \zeta$, $\theta_{\epsilon=0} = 0$ and $\phi_{\epsilon=0} = const$. \par 

A single resonant pulse is errant linearly $\theta_{\epsilon} = \theta(1+\epsilon)$, when for general composite rotation the particular forms of dependences on pulse area error $\epsilon$ of the three parameters are generally unknown and related to the structure of CPs, i.e. to the choice of pulse areas and composite phases for the certain number of pulses. 

Phasal $\zeta$ CPs belong to the case $\theta = 0$ and $\phi = const$. At least two CPs are required to obtain single phasal $\zeta$ pulse.

Note that derivation method presented in Subsec.~\ref{Subsec:uBBNBPB} does not care about rotation angle, geometric and relative phase stabilities. Here, we have deal with alternating CPs, which make the feature (robustness/sensitivity or both) of the pulse more powerful, sometimes called \textit{ultra}, at the expense of precision due to alternations (at the centre/ on the wings or both).

\subsection{Ultra-BB, ultra-NB and ultra-PB}\label{Subsec:uBBNBPB}
\subsubsection{Case of rotational \(\theta\) pulses}\label{Subsubsec:rot-pulses}
Let's maximise the population transfer area \eqref{eq-r-u-bn} at the whole-range of the error bandwidth, i.e. from $\epsilon = -1$ to $\epsilon = 1$ (ultrabroadband $\theta$ pulses) 
\be\label{eq-r-u-bn}
{\sum}_{b,n} \overset{\Delta}{=} \int_{-1}^{1} p(\epsilon) \,d{\epsilon} ,
\ee
or minimise it (ultranarrowband $\theta$ pulses). Here $p(\epsilon) = 1-|\mathcal{U}_{11}(\epsilon)|^2 = |\mathcal{U}_{12}(\epsilon)|^2 = \sin^2(\theta_{\epsilon}/2)$ is errant transition probability. 

In \eqref{eq-r-u-bn} $p(\epsilon=0)=p(\theta=\pi) = \sin^2{\theta/2}|_{\theta=\pi} = 1$, at the centre of bandwidth, is transition probability in QC: when pulse area error is zero, the qubit-state completely transfers from $\ket 0$ to $\ket 1$ due to $\pi$-rotation on the Bloch sphere. In PO this is mathematically equivalent (see Subsec. \ref{Sec:quantum-classical}) to the conversion of the polarisation state from $\ket L$ to $\ket R$ (or $\ket H$ to $\ket V$) due to $\pi$-rotation on the Poincar\'e sphere
\be\label{eq-conv-eff}
\int_{0}^{2\pi} I(\varphi{'}) \,d{\varphi{'}} = \int_{0}^{2\pi} |\mathcal{U}_{12}(\varphi {'})|^2 \,d{\epsilon}, 
\ee
and $I(\varphi{'})$ describes the conversion efficiency of the half-wave plate $I(\varphi{'} = \pi) = 1$. 

Let's maximise the population transfer area at the central half-range of the error bandwidth, i.e. from $\epsilon = -1/2$ to $\epsilon = 1/2$ (by minimising the area between phantom unit square and $p(\epsilon)$ profile line at the centre) and, simultaneously, minimise the p. t. area at the remaining part of the error bandwidth, i.e. at edge quarters $[-1, -1/2]$ and $[1/2, 1]$ (by minimising the area between profile $p(\epsilon)$ line and phantom $p(\epsilon) = 0$ line). Mathematically we minimise

\be\label{eq-r-u-p}
\begin{split}
& \left(1 - \int_{-1/2}^{1/2} p(\epsilon) \,d{\epsilon}\right) + \left(\int_{-1}^{-1/2} + \int_{1/2}^{1} p(\epsilon) \,d{\epsilon}\right) \overset{\Delta}{=} \\
& \quad \quad \quad \quad \quad \overset{\Delta}{=} \left(1 - {\sum}_{b}\right) + \left({\sum}_{n} \right),
\end{split}
\ee
where ${\sum}_{b}$ is a measure of broadness at the central-half, when ${\sum}_{n}$ is a measure of narrowness at the edge quarters. The corresponding pulses will have ultrasquare feature, and we name them ultrapassband $\theta$ pulses. Also, we denote ${\sum}_{p} \overset{\Delta}{=} {\sum}_{b} + {\sum}_{n} = \int_{-1}^{1} p(\epsilon) \,d{\epsilon}$ as the whole-range population transfer area in the case of ultrapassband $\theta$ pulses\footnote{Note that there is no inherent connection between the use of the same notations ${\sum}_{b}$ (and ${\sum}_{n}$) in the contexts of ultrapassband vs. ultrabroadband (and ultrapassband vs. ultranarrowband) other than the fact that they both serve as measures of broadness and narrowness, and refer to different ranges over the error bandwidth.}. In PO, equivalently, we minimise 
\be\label{eq-r-u-p-pol}
\begin{split}
& \left(1 - \int_{\pi/2}^{3\pi/2} I(\varphi') \,d{\varphi'}\right) + \left(\int_{0}^{\pi/2} + \int_{3\pi/2}^{2\pi} I(\varphi') \,d{\varphi'}\right) \overset{\Delta}{=} \\
& \quad \quad \quad \quad \quad \overset{\Delta}{=} \left(1 - {\sum}_{b}\right) + \left({\sum}_{n} \right).
\end{split}
\ee

Note that for rotational $\theta$ pulses, the target matrix is
\be\label{eq-theta-err}
\U_n = \left[ \begin{array}{cc} \cos(\theta/2) & -i \sin(\theta/2) \exp(i\phi) \\ -i \sin(\theta/2) \exp(-i\phi) & \cos(\theta/2) \end{array}\right],
\ee
and in the case of $\theta = \pi$ and $\phi = 0$ is equivalent to x-rotation on the Bloch sphere representing $\mathbf{R}_x(\pi) = e^{- i (\pi/2) \hat\sigma_x}$ rotation gate in the QC \cite{gevorgyan2021}. On the Poincar\'e sphere it maps to the Jones matrix for a half-waveplate $\textbf{J}_{0}(\pi)$ in the PO (see \eqref{J0half}).

\subsubsection{Case of phasal $\zeta$ pulses}\label{Subsubsec:phas-pulses}
Let's maximise the phase shifting area \eqref{eq-z-u-bn} at the whole-range of the error bandwidth, i.e. from $\epsilon = -1$ to $\epsilon = 1$ (ultrabroadband $\zeta$ pulses) 
\be\label{eq-z-u-bn}
\sum \overset{\Delta}{=} \int_{-1}^{1} z(\epsilon) \,d{\epsilon} .
\ee
Here the phase shifting $z(\epsilon) = \left(\mathcal{U}_{11}(\epsilon)-\mathcal{U}_{11}^{\ast}(\epsilon)\right)/(2i)$ is equal to the trace fidelity in our case $\zeta = \pi$
\be\label{eq-z-trace}
\begin{split}
& \mathcal{F}_{\text{T}} = \tfrac12 \text{Tr}\, [ \U_n (\epsilon) {\U_n}^\dagger ]  = \cos{\left(\frac{\zeta - \zeta_\epsilon}{2}\right)}_{\zeta = \pi} \cos{\left(\frac{\theta_\epsilon}{2}\right)} = \\
& = \sin{\left(\frac{\zeta_\epsilon}{2}\right)} \cos{\left(\frac{\theta_\epsilon}{2}\right)},
\end{split}
\ee
and the target matrix is
\be\label{eq-phase}
\U_n = \left[ \begin{array}{cc} \exp(- i\zeta/2) & 0 \\ 0 & \exp(i\zeta/2) \end{array}\right]_{\zeta=\pi} = \left[ \begin{array}{cc} - i & 0 \\ 0 & i \end{array}\right],
\ee
which corresponds to the z-rotation on the Bloch sphere acting as $\mathbf{Z}$ phase gate in the QC (see \cite{gevorgyan2024}). On the Poincar\'e sphere it matches with the Jones matrix for a polarisation rotator $\textbf{J}_{\pi}$ in the PO (see \eqref{Jt}).

\section{Ultrabroadband rotational $\theta=\pi$ pulses}\label{Sec:ultrabb}

The most convenient way to construct ultrabroadband rotational $\pi$ pulses is the symmetric design consisting of nominal $\pi$ pulses

\be\label{ultrabb_theta}
\pi_{\phi_1} \pi_{\phi_2} \ldots \pi_{\phi_{k/2}} \pi_{\phi_{k/2+1}} \pi_{\phi_{k/2}} \ldots \pi_{\phi_2} \pi_{\phi_1},
\ee
where $k = N-1$ is the number of inflection points in the errant transition probability vs the pulse area error plot (the number of alternations of the profile). Since the relative constituent phases play a significant role in the calculation, the first and the last phases can be taken as zero $\phi_1=0$.

\begin{figure}[t]
% \bt{r}
% \centerline{\includegraphics[width=0.947\columnwidth]{Figures/fig8a.eps}}
% \et
\centering
\includegraphics[width=0.947\columnwidth]{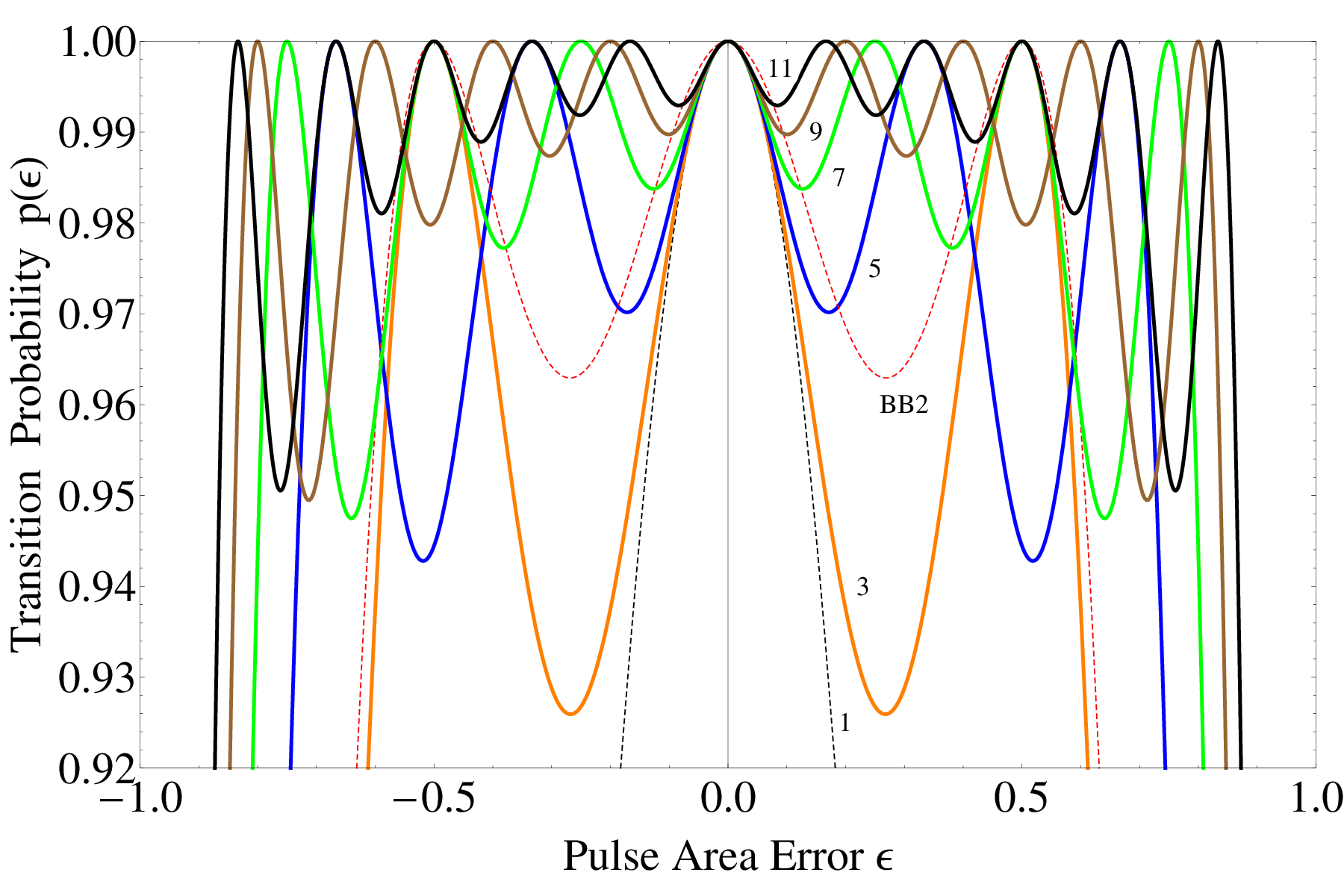}
\captionsetup{justification=raggedright,singlelinecheck=false} % align caption left
\caption{
Transition probability $p(\epsilon)$ of ultrabroadband rotational $\pi$ pulses.
The numbers $N$ on the curves refer to composite sequences UB$N$ (named Bat$N$) listed in Table \ref{Table:ultrarobust-theta}.
As noted above, the curves have $k = N-1$ alternations on the top of the plot, unlike the BB2 sequence, which has 2 alternations, so it's worse than our five-$\pi$ Bat5.
}
\label{fig:UB}
\end{figure}

Ultrabroadband rotational $\theta = \pi$ CPs, derived by using the method Subsec.~\ref{Subsec:uBBNBPB}, have maximum state transfer area for the certain number of CPs, hence are unique. For example, five-pulse sequence UB5 with 4 alternations is better than the well-known BB2 sequence with 2 alternations. We have derived up to eleven sequences, which increase the broadness range of the original rotational sequence (a single pulse) more than four times (from 20.5\% to 87.7\%), and the transition probability area is increased by 83.(3)\% by the eleven-$\pi$ UB11 sequence. Composite phases for the ultrabroadband rotational pulses are shown in the Table~\ref{Table:ultrarobust-theta}, and the transition probability is plotted in Figure~\ref{fig:UB}. For comparison, our five-$\pi$ UB5 CP sequence has the transition probability area equal to $\frac{5}{3}=1.(6)$, which is smaller than the area of about $\frac{1}{8} (11+\sqrt{2}) \approx 1.552$ of the well-known five-$\pi$ BB2 sequence, i.e. by about 0.115. Error robustness range of UB5 is equal to 75.2\% and is broader than the range of 64.4\% of BB2 sequence. 

\section{Ultranarrowband rotational $\theta=\pi$ pulses}\label{Sec:ultranb}

Since NB pulses are asymmetric in composite phases, to construct ultranarrowband rotational $\pi$ pulses we choose the antisymmetric design consisting of nominal $\pi$ pulses

\be\label{ultranb_theta}
\pi_{\phi_1} \pi_{\phi_2} \ldots \pi_{\phi_{k/2}} \pi_{\phi_{k/2+1}} \pi_{-\phi_{k/2}} \ldots \pi_{-\phi_2} \pi_{-\phi_1},
\ee
where $k = N-1$ is the number of inflection points in the errant transition probability vs the pulse are error plot (the number of alternations of the profile). For convenience, the middle phases can be taken as ${\phi}_{k/2+1}=\pi$.

\begin{figure}[t]
\bt{r}
\centerline{\includegraphics[width=0.947\columnwidth]{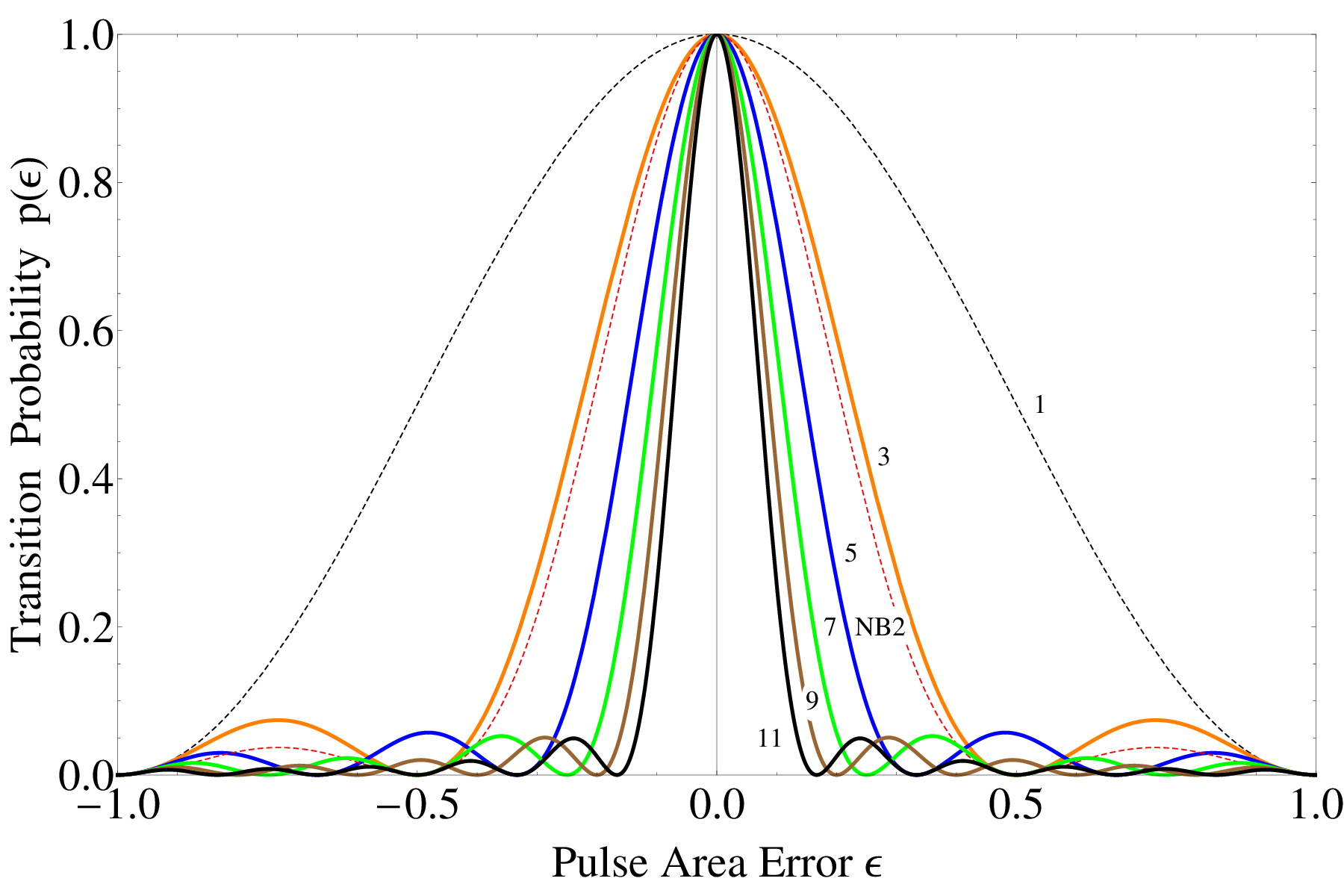}}
\et
\captionsetup{justification=raggedright,singlelinecheck=false} % align caption left
\caption{
Transition probability $p(\epsilon)$ of ultranarrowband rotational $\pi$ pulses.
The numbers $N$ on the curves refer to composite sequences UN$N$ (named Snake$N$) listed in Table \ref{Table:ultrasensitive-theta}.
As noted above, the curves have $k = N-1$ alternations on the bottom of the plot, unlike the NB2 sequence, which has 2 alternations, so it's worse than our five-$\pi$ Snake5.
}
\label{fig:UN}
\end{figure}

Ultranarrowband rotational $\theta = \pi$ CPs, derived by using the method Subsec.~\ref{Subsec:uBBNBPB}, have minimum state transfer area for the certain number of CPs, hence are unique. For example, five-pulse sequence UN5 with 4 alternations is better than the well-known NB2 sequence with 2 alternations. We have derived up to eleven sequences, which decrease the narrowness range at 50\% of probability, viz. full width at half maximum (FWHM), of the original rotational sequence (a single pulse) about 6.75 times (from 50\% to 7.4\%), and the transition probability area is decreased by 83.(3)\% by the eleven-$\pi$ UN11 sequence. Composite phases for the ultranarrowband rotational pulses are shown in the Table~\ref{Table:ultrasensitive-theta}, and the transition probability is plotted in Figure~\ref{fig:UN}. For comparison, our five-$\pi$ UN5 CP sequence has the transition probability area equal to $\frac{1}{3}=0.(3)$, which is smaller than the area of about $\frac{1}{8} (5-\sqrt{2}) \approx 0.448$ of the well-known five-$\pi$ NB2 sequence, i.e. by about 0.115. Error sensitivity range of UN5 at FWHM is equal to 14.9\% and is narrower than the FWHM range of 20.8\% of NB2 sequence.

\section{Ultrapassband rotational $\theta=\pi$ pulses}\label{Sec:ultrapb}

Following the structure of Wimperis' passband pulses PB1 and PB2, we construct ultrapassband rotational $\pi$ pulses according to the next design 

\be\label{ultrapb_theta}
\pi_{\phi_1} (2\pi)_{\phi_2} (2\pi)_{\phi_3} \ldots (2\pi)_{\phi_{k/2+1}},
\ee
where $k = 2(N-1)$ is the number of inflection points in the errant transition probability vs the pulse are error plot (the number of alternations of the profile). Inflection points at the bottom and at the top are equal $k_n = k_b = k/2 = N-1$. 

Ultrapassband rotational $\theta = \pi$ CPs, derived by using the method Subsec.~\ref{Subsec:uBBNBPB}, have maximum state transfer area at the centre $\epsilon = [-1/2, 1/2]$ and minimum state transfer area on the wings $\epsilon = [-1, -1/2]$ and $\epsilon = [1/2, 1]$ for the certain number of CPs, hence are unique. More specifically, this is the special case of ultrapassband class, considering that, in general, the class may consist of pulses with non-necessarily equal properties of broadness and narrowness, namely, ultrarectangular pulses. In our case, ultrasquare pulse profile is symmetrical with respect to the top and bottom. This particularly means that every pair of inflection points at the top have corresponding pair of inflection points at the bottom, and they are in the same distance from $\epsilon = 0$ for the top and $\epsilon = \pm 1$ for the bottom, phantom vertical lines. Moreover, their transition probability values are at the same distance from $p(\epsilon) = 1$ for the top and $p(\epsilon) = 0$ for the bottom, phantom horizontal lines. This general symmetry can be mathematically expressed by the relations $\epsilon (p) = 1 - \epsilon (1 - p)$ and $p(\epsilon) + p(1-\epsilon) = 1$ for any point $(\epsilon, p)$ of the profile, where for the latter transition probability is considered as a function of the error ($p(\epsilon)$), and vice versa for the former ($\epsilon(p)$). Although, ultrasquare and ultrarectangular pulses may differ in broadness-narrowness ratio, but they both have a property of high rectangularity --- an indistinguishable feature for passband pulses. We propose to use $\varkappa = \frac{\Delta(p)}{\Delta(\epsilon)} = \frac{1 - 2\alpha}{\Delta(\alpha)}$ as the measure of rectangularity of passband CPs, where $\Delta(p) = p_b - p_n$, and $p_b$ and $p_n$ are the probability transition benchmarks at the top and at the bottom, respectively. In our case, phantom benchmark lines are taken equally $\alpha \overset{\Delta}{=} p_n$ distant from the top $p=1$ and from the bottom $p=0$, respectively. Hence, the corresponding distance in transition probability is $\Delta(p) = 1 - 2\alpha$ and corresponding distance in error is $\Delta(\epsilon) = \lvert \epsilon_n \rvert - \lvert \epsilon_b \rvert =  \epsilon(\alpha) - \epsilon(1-\alpha) \overset{\Delta}{=} \Delta(\alpha)$ (Note that we consider just one side of the profile: positive or negative side of error). Due to the performance of our pulses, we target applications for which $\alpha$ equal to $10^{-1}$ is sufficient, and rectangularity measure $\Delta \overset{\Delta}{=} \Delta(0.1)$ is the difference between absolute errors at low (10\%) and high (90\%) transition probabilities. Since the slope coefficient (is approximated by a straight line $\tan{\beta} \simeq \varkappa$) is inversely proportional to $\Delta$, hence, smaller $\Delta$, higher the rectangularity of the transition probability profile.    

As an illustration, consider the three-pulse sequence UPB3 featuring four alternations. Despite exhibiting slightly inferior performance in metrics such as ${\sum}_{b}$, ${\sum}_{n}$, and error correction range, UPB3 surpasses PB2 notably in terms of rectangularity (1.065 times more), requiring approximately  fewer pulses, specifically two fewer $2\pi$ pulses, and demonstrating an execution time that is approximately 1.8 times faster.

\begin{figure}[htbp]
  \centering
  \begin{subfigure}[b]{1\columnwidth}
    \centering
    \includegraphics[width=0.947\columnwidth]{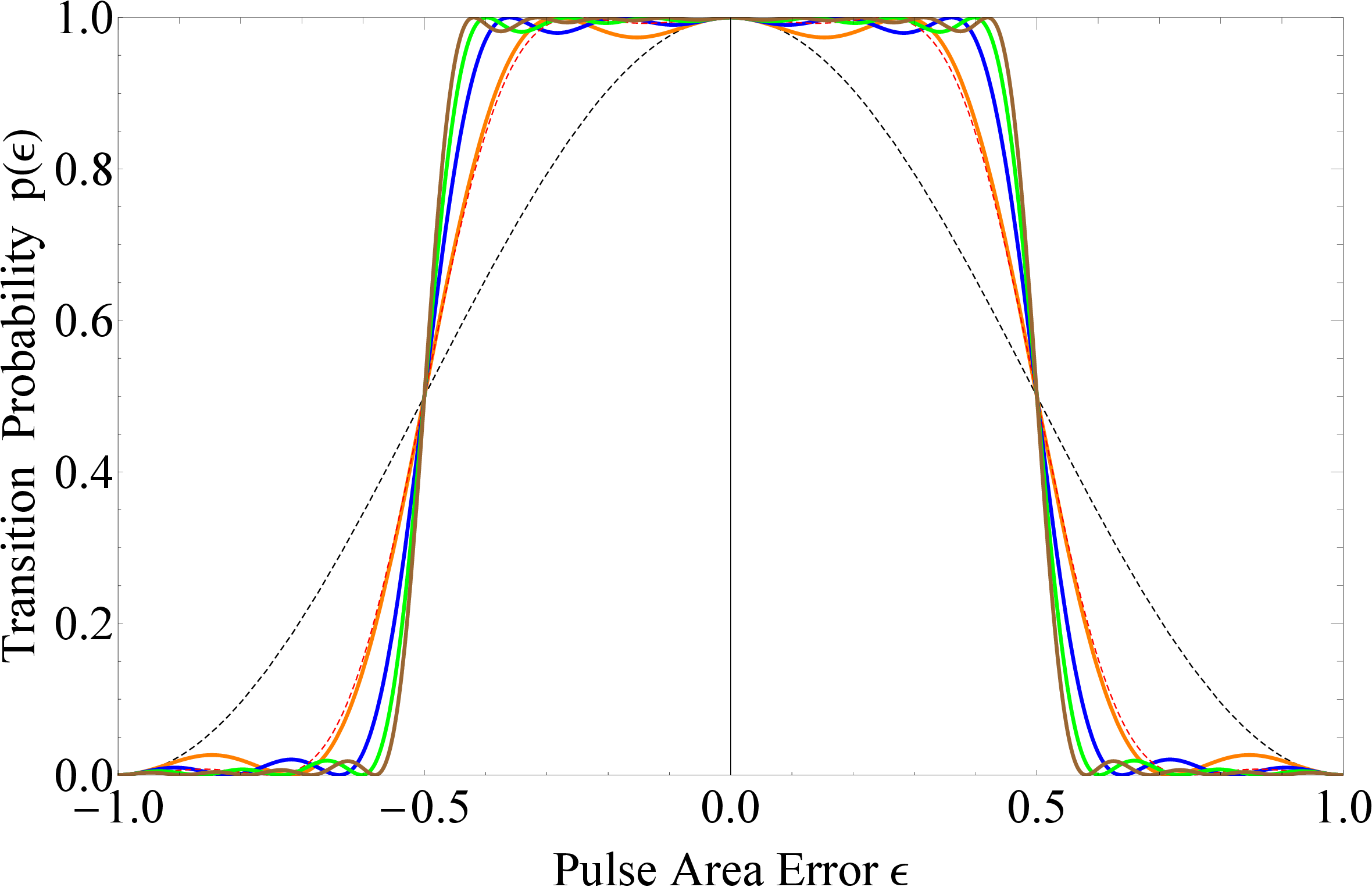}    
    \caption{Figure as a whole.}
    \label{fig:sub1}
  \end{subfigure}
  
  \begin{subfigure}[b]{1\columnwidth}
    \centering
    \includegraphics[width=0.947\columnwidth]{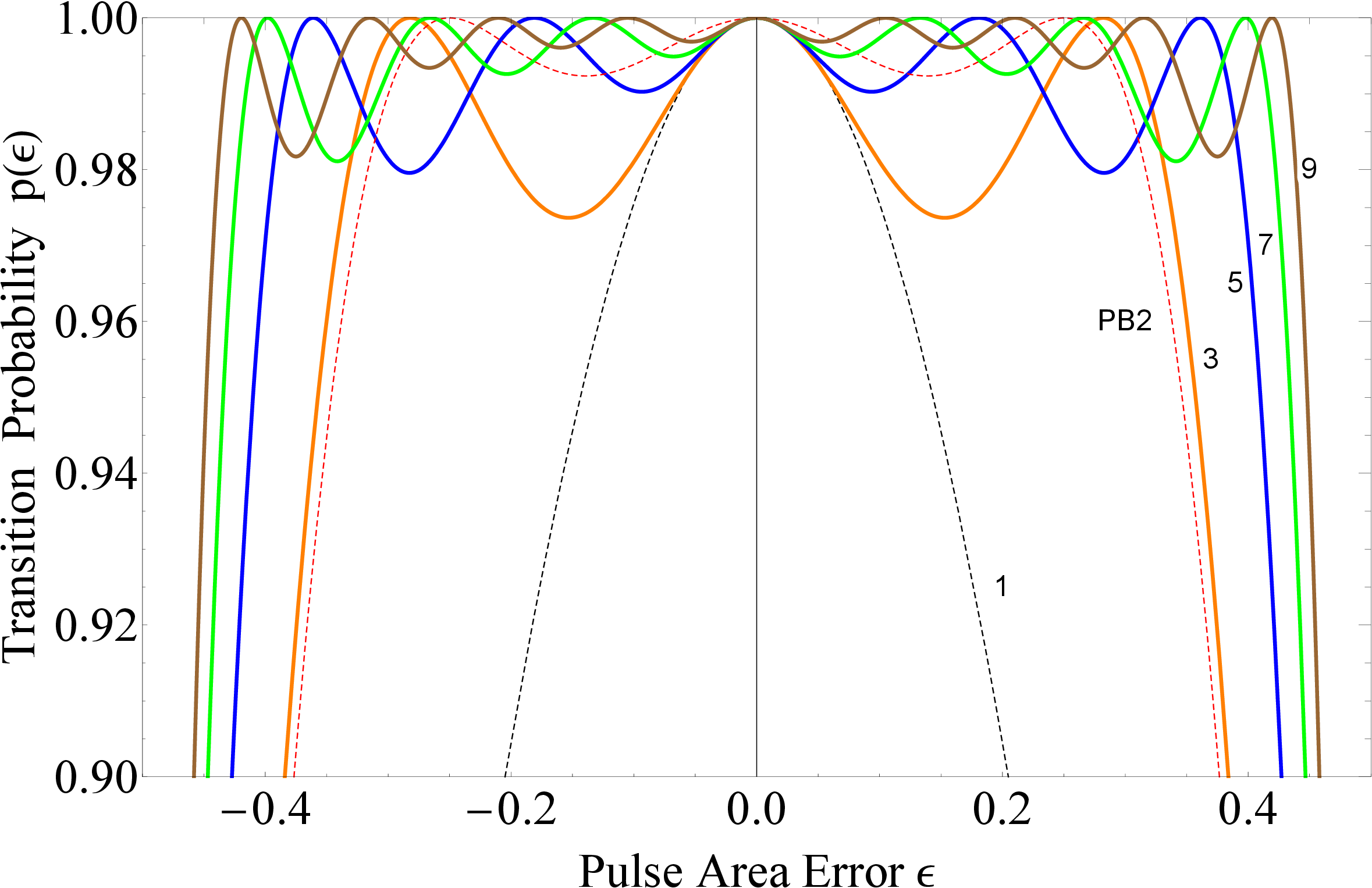}    
    \caption{Figure at the top and at the central part of the pulse area error bandwidth.}
    \label{fig:sub2}
  \end{subfigure}
   
  \begin{subfigure}[b]{1\columnwidth}
    \centering
    \includegraphics[width=0.947\columnwidth]{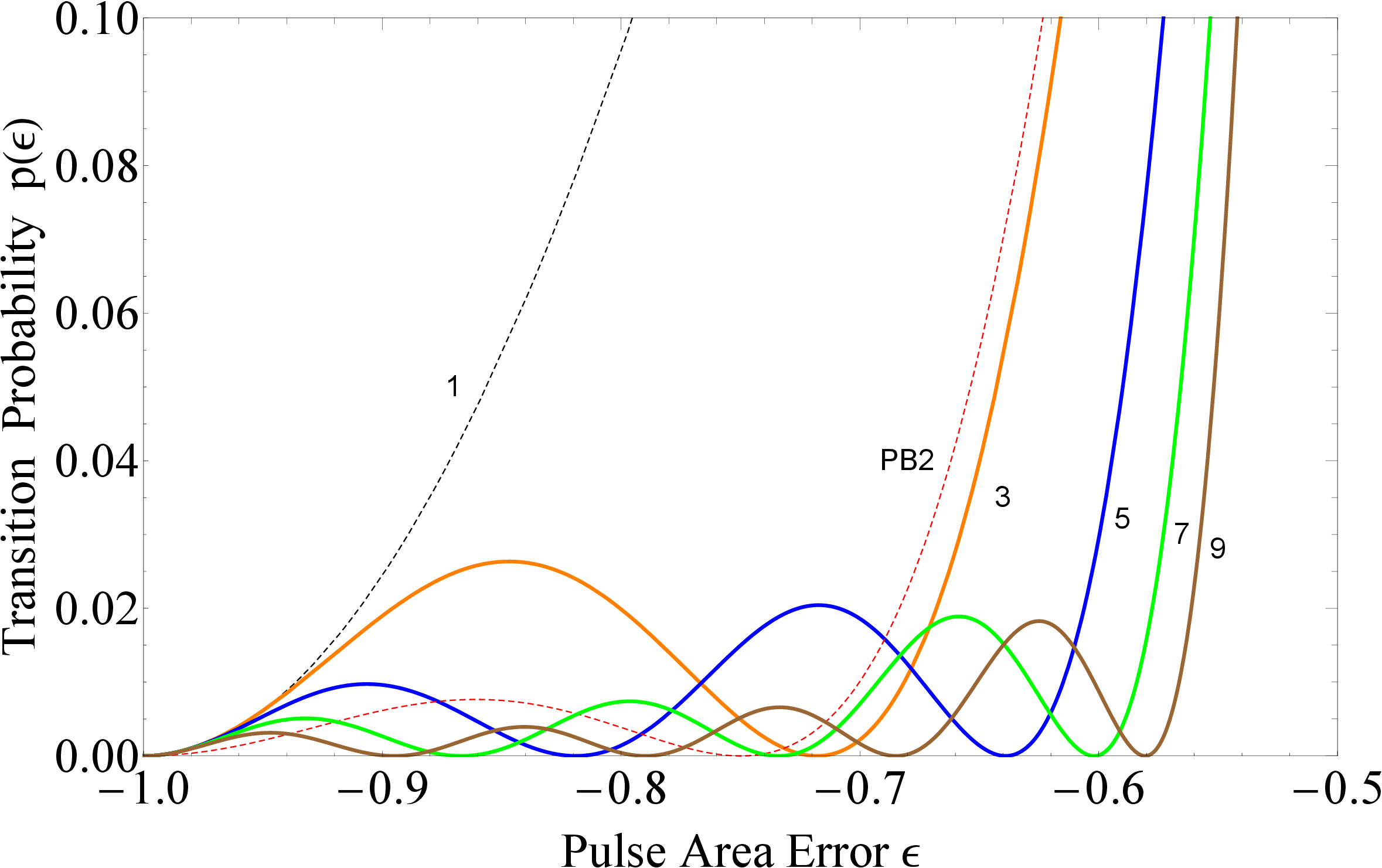}    
    \caption{Figure at the bottom and at the negative edge of the pulse area error bandwidth.}
    \label{fig:sub3}
  \end{subfigure}  \captionsetup{justification=raggedright,singlelinecheck=false} % align caption left
  \caption{Transition probability $p(\epsilon)$ of ultrapassband rotational $\pi$ pulses. The numbers $N$ on the curves refer to composite sequences UPB$N$ (named Octopus$N$) listed in Table \ref{Table:ultrasquare-theta}. As noted above, the curves have $k_{b,n} = N-1$ alternations on the top and on the bottom of the plot, respectively (total of $k = 2 k_{b,n} = 2 (N-1)$), unlike the PB2 sequence, which has 2 alternations, so it's worse than our nine-$\pi$ Octopus5.}
  \label{fig:UPB}
\end{figure}

As an illustration, consider the three-pulse sequence UPB3 featuring four alternations. Despite exhibiting slightly inferior performance in metrics such as ${\sum}_{b}$, ${\sum}_{n}$, and error correction range, UPB3 surpasses the well-known PB2 notably in terms of rectangularity (is roughly 1.065 times more squared), required number of pulses (requires two fewer $2\pi$ pulses) and execution time (1.8 times faster). We have derived up to nine sequences, which, compared to the original rotational sequence (a single pulse), increase the broadness range of  close to an ideal (from 20.5\% to 45.8\%) (Note that an ideal is 50\% in the case of an ideal square pulse.), increase the central transition probability area by roughly by 18.7\% (from $\frac{1}{2} + \frac{1}{\pi} \approx 0.818$ to $0.971$), approximately covering  the central transition probability area of the central square, i.e. 97.1\% (Note that ${\sum}_{p} = {\sum}_{b} + {\sum}_{n}$ is always 1 for all square passband pulses, and, interestingly, a single pulse has this property too.), and increase the rectangularity measure roughly by seven times (from 1.36 to 9.54) (It is noteworthy that a single pulse possesses an approximate slope exceeding $45^\circ$, as evidenced by a rectangularity measure exceeding 1.). Composite phases for the ultrapassband rotational pulses are shown in the Table~\ref{Table:ultrasquare-theta}, and the transition probability is plotted in Figure~\ref{fig:UPB}.

\section{Ultrabroadband phasal $\zeta=\pi$ pulses}\label{Sec:ultrabbph}

As usual \cite{gevorgyan2024}, we construct ultrabroadband phasal $\pi$ pulses with asymmetric design consisting of nominal $\pi$ pulses

\be\label{ultrabb_phase}
\pi_{\phi_1} \pi_{\phi_2} \ldots \pi_{\phi_{k/2+1}} \cdot \pi_{\phi_{1} + \frac{1}{2}\pi} \pi_{\phi_{2} + \frac{1}{2}\pi} \ldots \pi_{\phi_{k/2+1} + \frac{1}{2}\pi} ,
\ee
where $k = N-2$ is the number of inflection points in the trace fidelity vs the pulse are error plot (the number of alternations of the plot). Careful analysis shows that the first few phases can be taken as zero in the calculation (cf. Table~\ref{Table:ultrarobust-phase}).

\begin{figure}[t]
\bt{r}
\centerline{\includegraphics[width=0.947\columnwidth]{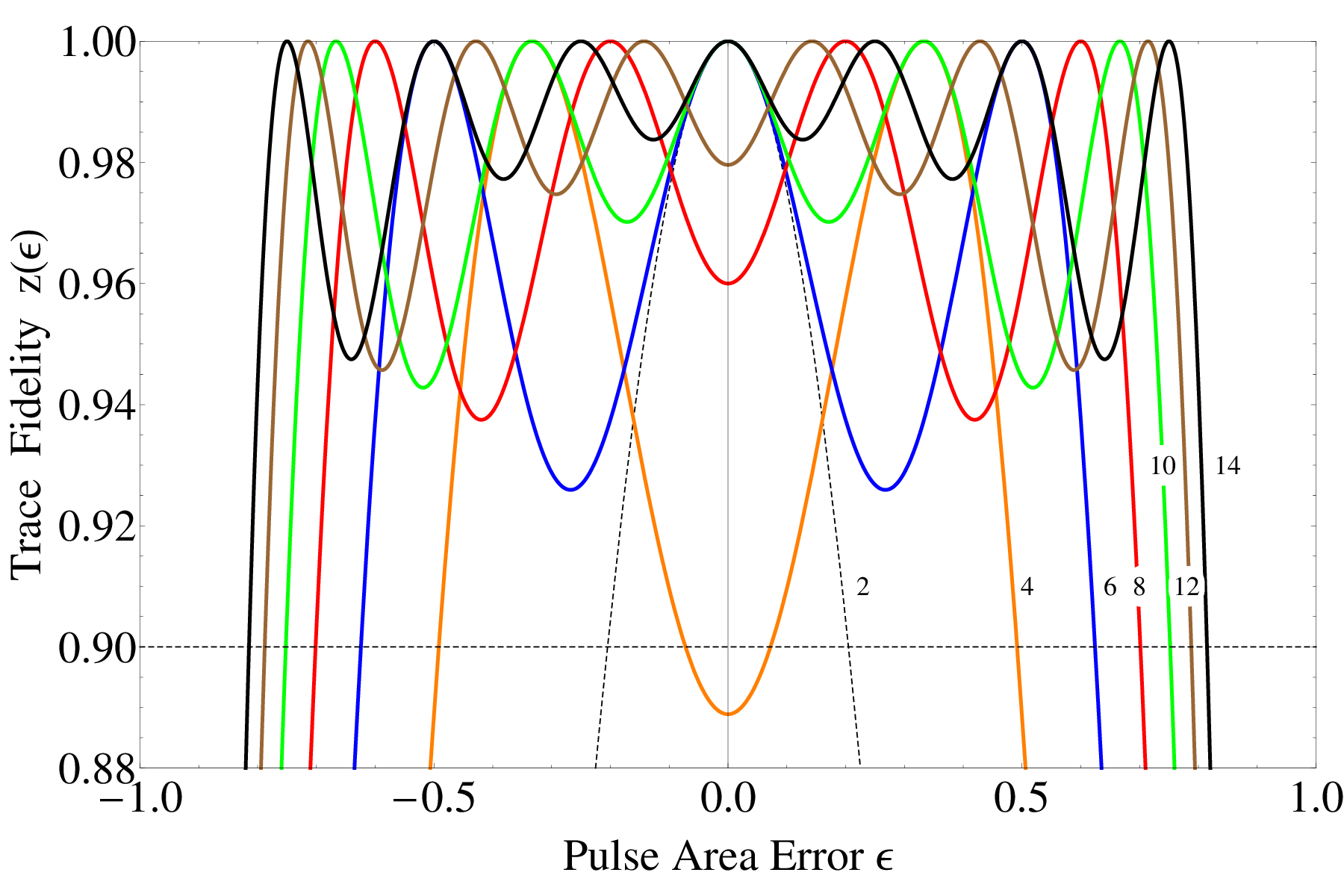}}
\et
\captionsetup{justification=raggedright,singlelinecheck=false} % align caption left
\caption{
Trace fidelity $z(\epsilon)$ of ultrabroadband phasal $\pi$ pulses.
The numbers $N$ on the curves refer to composite sequences UBPh$N$ listed in Table \ref{Table:ultrarobust-phase}.
As noted above, the curves have $k = N-2$ alternations on the top of the plot\footnote{Four, eight and twelve phasal sequences are below 100\% fidelity at the centre (errorless case). Moreover, the trace fidelity is slightly less than 90\% in the case of the UBPh4 sequence. Note that in some applications where ultrahigh precision is not a mandatory criterion, these violations are minor deviations from the requirements, and these composite sequences can be applied.}.
}
\label{fig:UBPh}
\end{figure}

Ultrabroadband phasal $\zeta = \pi$ CPs, derived by using the method Subsec.~\ref{Subsec:uBBNBPB}, have maximum trace fidelity area for the certain number of CPs, hence are unique. We have derived up to fourteen sequences, which increase the broadness range of the original phasal sequence (two pulses) about four times (from 20.5\% to 81.5\%), and the trace fidelity area is increased by the 75\% by the fourteen-$\pi$ UBPh14 sequence. Composite phases for the ultrabroadband phasal pulses are shown in the Table~\ref{Table:ultrarobust-phase}, and the trace fidelity is plotted in Figure~\ref{fig:UBPh}\footnote{Four, eight and twelve phasal sequences are below 100\% fidelity at the centre (errorless case). Moreover, the trace fidelity is slightly less than 90\% in the case of the UBPh4 sequence. Note that in some applications where ultrahigh precision is not a mandatory criterion, these violations are minor deviations from the requirements, and these CP sequences can be applied.}.

%%%%%%%%%%%%%%%%%%%%%%%%%%%%%%%%%%%%%%%%%%%%%%%%%%%%%%%%%%%%%%%%%%%%%%%%%%%%%%%%%%%%%%%%%%%%%%%%%%%%%%%%%%%%%%%%%%%%%%%%%%%%%%%%%%%%%%%%%
%%%%%%%%%%%%%%%%%%%%%%%%%%%%%%%%%%%%%%%%%%%%%%%%%%%%%%%%%%%%%%%%%%%%%%%%%%%%%%%%%%%%%%%%%%%%%%%%%%%%%%%%%%%%%%%%%%%%%%%%%%%%%%%%%%%%%%%%%
\section{Comments and Conclusions\label{Sec:concl}}
%%%%%%%%%%%%%%%%%%%%%%%%%%%%%%%%%%%%%%%%%%%%%%%%%%%%%%%%%%%%%%%%%%%%%%%%%%%%%%%%%%%%%%%%%%%%%%%%%%%%%%%%%%%%%%%%%%%%%%%%%%%%%%%%%%%%%%%%%
%%%%%%%%%%%%%%%%%%%%%%%%%%%%%%%%%%%%%%%%%%%%%%%%%%%%%%%%%%%%%%%%%%%%%%%%%%%%%%%%%%%%%%%%%%%%%%%%%%%%%%%%%%%%%%%%%%%%%%%%%%%%%%%%%%%%%%%%%
We presented a number of CP sequences consisting of $\pi$ pulses for transition of the quantum state from $\ket{0}$ to $\ket{1}$ in ultrarobust and ultrasensitive manners, according to the pulse area deviation $\epsilon$. Using quantum-classical analogy, we presented a number of sequences of half-wave plates for conversion of the polarisation state from $\ket{H}$ to $\ket{V}$ or from $\ket{L}$ to $\ket{R}$ in ultrabroadband and ultranarrowband ways, according to the phase-shift (retardation) deviation $\varphi\prime - \varphi = \varphi\prime - \pi$. Our UB5 pulse already outperforms the well-known BB2 pulse in terms of broadness, e.g. UB5 maintains 90\% of transition probability (or conversion efficiency) over the broadness (error-correction or retardation deviation) range spanning a width of roughly $1.504\pi$ from the whole $2\pi$, approximately by 17\% larger than roughly $1.288\pi$, the width of BB2. Our longest UB11 pulse covers approximately 88\% of the whole width for the same benchmark. Our UN5 pulse already outperforms the well-known NB2 pulse in terms of narrowness, e.g. FWHM of UN5 is roughly $0.298\pi$, approximately $1.4$ times narrower than FWHM of NB2. Our longest UNB11 pulse covers approximately 21\% of the whole width for the same benchmark.   

Furthermore, using the similar derivation approach of CPs, we theoretically design ultrarobust Z quantum gate via a number of CP sequences consisting of $\pi$ pulses, and equivalently ultrabroadband polarisation $\pi$ rotator. Our longest UBPh14 pulse maintains 90\% of trace fidelity over a broadness range of roughly $1.63\pi$. 

With the choice of the pulse area structure (or the phase-shift structure in PO) of the CP (or combination of wave-plates in PO), one can apply the method of derivation to obtain arbitrary transition (or arbitrary conversion in PO) from the given quantum (or polarisation) state to the arbitrary quantum (or polarisation) state in ultrabroadband and ultranarrowband manners. Certainly, achieving superposition state $\frac{1}{\sqrt{2}}\left(\ket{0} \pm i \ket{1}\right)$\footnote{An ideal $\pi/2$ rotation on the Bloch sphere from $\ket{0}$ or $\ket{1}$ initial states presents these states. Actual superposition states $\frac{1}{\sqrt{2}}\left(\ket{0} \pm \ket{1}\right)$ can be obtained with the same fashion targeting $\pi/2$ rotation with phase $\phi = \pi/2$.} (or left-right circular polarisation bases $\ket{L}$, $\ket{R}$) is of interest. 

Results are promising for applications in NMR, QS and, especially PO, where the property of robustness/broadness or selectivity/narrowness is more important and ultrahigh-precision is not obligatory as in QC. In this sense, we acknowledge also the future applications that are not on demand due to the absence of the method.  

Worth to mention that the article is the advanced version of the similar idea in the 2022's conference paper \cite{gevorgyan2022}. Note that in the Figures~\ref{fig:UB},~\ref{fig:UN},~\ref{fig:UPB} in the vertical axis, the transition probability $p(\epsilon)$ can be changed to the conversion efficiency $I(\varphi)$, and correspondingly in the horizontal axis, the pulse area error $\epsilon$ to the phase shift $\varphi$, and in that case, $\epsilon = 0.0$ will correspond to $\varphi = \pi$, $\epsilon = 0.5 \rightarrow \varphi = 1.5\pi$, $\epsilon = 1.0 \rightarrow \varphi = 2.0\pi$, $\epsilon = -0.5 \rightarrow \varphi = 0.5\pi$, $\epsilon = -1.0 \rightarrow \varphi = 0.0\pi$.

%========================

%%%%%%%%%%%%%%%%%%%%%%%%%%%%%%%%%%%%%%%%%%%%%%%%%%%%%%%%%%%%%%%%%%%%%%%%%%%%%%%%%%%%%%%%%%%%%%%%%%%%%%%%%%%%%%%%%%%%%%%%%%%%%%%%%%%%%%%%%%%%%%%%%%%%%%%%%%%%%%%%

\acknowledgments
HLG acknowledges support from the EU Horizon-2020 ITN project LIMQUET \textit{Light-Matter Interfaces for Quantum Enhanced Technologies} (Contract No. 765075), and also from the Higher Education and Science Committee of Armenia in the frames of the research project 21AG-1C038 on \textit{Methods of Information Theory in Statistical Physics and Data Science}. 
%\newpage
\appendix

\section{Appendix}

%***************************************************************
%***************************************************************

%***************************************************************
%***************************************************************

%***************************************************************
%***************************************************************

%***************************************************************
%***************************************************************

%T%T%T%T%T%T%T%T%T%T%T%T%T%T%T%T%T%T%T%T%T%T%T%T%T%T%T%T%T%T%T%T%T%T%T%T%T%T%T%T%T%T%T%T%T
\begin{table*}[tbph]
\begin{tabular}{|c|c|c|c|c|c|}
\hline
{\bfseries Name } & {\bfseries Pulses } & {\bfseries Number of} & {\bfseries $\sum_b$} & {\bfseries Phases } & {\bfseries Transition probability} \\
& & {\bfseries alternations $k$} & {\bfseries {cf. \eqref{eq-r-u-bn}}} & {\bfseries $\phi_1,\phi_2,\phi_3, \ldots, \phi_{k/2}, \phi_{k/2+1}, \phi_{k/2}, \ldots, \phi_{3}, \phi_{2}, \phi_{1}$} & {\bfseries $p(\epsilon)=90\%$} \\
 & & {\bfseries (inflection points)} & & {\bfseries (in units $\pi$) (according to \eqref{ultrabb_theta})} & {\bfseries error correction range}\\
\hline
single & 1 & 0 & $1$ & $0$ & $[0.795\pi, 1.205\pi]$\\
\rowcolor{lime}%
Bat3 & 3 & 2 & $1.5$ & $0, \frac12$ & $[0.376\pi, 1.624\pi]$\\
Bat5 & 5 & 4 & $1.(6)$ & $0, 0.5825, 0.3737$ & $[0.248\pi, 1.752\pi]$\\
\rowcolor{lime}%
Bat7 & 7 & 6 & $1.75$ & $0, 0.6230, 0.4918, 0.7558$ & $[0.185\pi, 1.815\pi]$\\
Bat9 & 9 & 8 & $1.8$ & $0, 0.6490, 0.5514, 0.8458, 0.6774$ & $[0.148\pi, 1.852\pi]$\\
\rowcolor{lime}%
Bat11 & 11 & 10 & $1.8(3)$ & $0, 0.6677, 0.5886, 0.9044, 0.7786, 0.9663$ & $[0.123\pi, 1.877\pi]$\\
\hline
BB2 & 5 & 2 & $\approx 1.552$ & $0, \frac12, \frac74, \frac74, \frac12$ & $[0.356\pi, 1.644\pi]$\\
\hline
\end{tabular}
\captionsetup{justification=raggedright,singlelinecheck=false} % align caption left
\caption{
Phases of symmetric, altering, ultrabroadband composite rotational $\theta = \pi$ sequences of $N=k+1$ nominal $\pi$ pulses, which produce ultrarobust population transfer in the ultrabroadband pulse area error correction range.
The last column gives the high-transition probability range $[\pi (1-\epsilon_0), \pi (1+\epsilon_0)]$ of pulse area error compensation wherein the errant transition probability is above the value $0.9$. The term Bat$N$ is attributed to UB$N$ composite pulses because of their resemblance in appearance.}
\label{Table:ultrarobust-theta}
\end{table*}
%T%T%T%T%T%T%T%T%T%T%T%T%T%T%T%T%T%T%T%T%T%T%T%T%T%T%T%T%T%T%T%T%T%T%T%T%T%T%T%T%T%T%T%T%T

%T%T%T%T%T%T%T%T%T%T%T%T%T%T%T%T%T%T%T%T%T%T%T%T%T%T%T%T%T%T%T%T%T%T%T%T%T%T%T%T%T%T%T%T%T
\begin{table*}[tbph]
\begin{tabular}{|c|c|c|c|c|c|}
\hline
{\bfseries Name } & {\bfseries Pulses } & {\bfseries Number of} & {\bfseries $\sum_n$} & {\bfseries Phases } & {\bfseries FWHM: $p(\epsilon)=50\%$} \\
 & & {\bfseries alternations $k$} & {\bfseries {cf. \eqref{eq-r-u-bn}}} & {\bfseries $\phi_1,\phi_2,\phi_3, \ldots, \phi_{k/2}, \phi_{k/2+1}, -\phi_{k/2}, \ldots, -\phi_{3}, -\phi_{2}, -\phi_{1}$} & {\bfseries transition probability} \\
 & & {\bfseries (inflection points)} & & {\bfseries (in units $\pi$) (according to \eqref{ultranb_theta})} & {\bfseries error sensitivity range}\\
\hline
single & 1 & 0 & $1$ & $0$ & $[0.5\pi, 1.5\pi]$\\
\rowcolor{pink}%
Snake3 & 3 & 2 & $0.5$ & $\frac12, 1$ & $[0.772\pi, 1.228\pi]$\\
Snake5 & 5 & 4 & $0.(3)$ & $0.5896, 0.4104, 1$ & $[0.851\pi, 1.149\pi]$\\
\rowcolor{pink}%
Snake7 & 7 & 6 & $0.25$ & $0.5193, 0.6121, 0.3671, 1$ & $[0.889\pi, 1.111\pi]$\\
Snake9 & 9 & 8 & $0.2$ & $0.5451, 0.4880, 0.6235, 0.3340, 1$ & $[0.911\pi, 1.089\pi]$\\
\rowcolor{pink}%
Snake11 & 11 & 10 & $0.1(6)$ & $0.5173, 0.5562, 0.4690, 0.6312, 0.3209, 1$ & $[0.926\pi, 1.074\pi]$\\
\hline
NB2 & 5 & 2 & $\approx 0.448$ & $0, \frac12, \frac54, \frac54, \frac12$ & $[0.792\pi, 1.208\pi]$\\
\hline
\end{tabular}
\captionsetup{justification=raggedright,singlelinecheck=false} % align caption left
\caption{
Phases of asymmetric, altering, ultranarrowband composite rotational $\theta = \pi$ sequences of $N=k+1$ nominal $\pi$ pulses, which produce ultrasensitive population transfer in the ultranarrowband pulse area error sensitivity range.
The last column gives the full width at half maximum range $[\pi (1-\epsilon_0), \pi (1+\epsilon_0)]$ of pulse area error sensitivity wherein the errant transition probability is above the value $0.5$. Note that the full population transfer occurs at the centre for zero pulse area error $p(\epsilon=0)=1$. The term Snake$N$ is attributed to UN$N$ composite pulses because of their resemblance in appearance.}
\label{Table:ultrasensitive-theta}
\end{table*}
%T%T%T%T%T%T%T%T%T%T%T%T%T%T%T%T%T%T%T%T%T%T%T%T%T%T%T%T%T%T%T%T%T%T%T%T%T%T%T%T%T%T%T%T%T

%T%T%T%T%T%T%T%T%T%T%T%T%T%T%T%T%T%T%T%T%T%T%T%T%T%T%T%T%T%T%T%T%T%T%T%T%T%T%T%T%T%T%T%T%T
\begin{table*}[tbph]
\begin{tabular}{|c|c|c|c|c|c|c|c|}
\hline
{\bfseries Notation } & {\bfseries Pulses / } & {\bfseries $k_b = k_n = k/2$}  
& {\bfseries $\sum_b$} & {\bfseries $\sum_n$} & {\bfseries Phases } & {\bfseries T. p.} & {\bfseries Rectan-}\\
 & {\bfseries Execution } & {\bfseries at the top /} & {\bfseries {cf. \eqref{eq-r-u-p}}} & {\bfseries {cf. \eqref{eq-r-u-p}}} & {\bfseries $\phi_1,\phi_2,\phi_3, \ldots, \phi_{k/2+1}$} & {\bfseries $p(\epsilon)=90\%$} & {\bfseries gularity} \\
 & {\bfseries time } & {\bfseries bottom} & & & {\bfseries (in units $\pi$) (according to \eqref{ultrapb_theta})} & {\bfseries e. c. range} & {\bfseries $\varkappa$}\\
\hline
single & 1 / $\pi$ & 0 & $0.818$ & $0.182$ & $0$ & $[0.795\pi, 1.205\pi]$ & 1.36 \\
\rowcolor{Apricot}%
Octopus3 & 3 / $5\pi$ & 2 & $0.921$ & $0.079$ & $0, 0.4691, 1.1808$ & $[0.616\pi, 1.384\pi]$ & 3.45 \\
%0.4691212055782626`, 1.1808304682159256`
Octopus5a & 5 / $9\pi$ & 4 & $0.950$ & $0.050$ & $0, 0.2882, 1.8507, 1.0435, 1.2262$ & $[0.573\pi, 1.427 \pi]$ & 5.49 \\
Octopus5b  & & & & & $0, 0.5662, 1.0608, 1.2123, 1.9112$ & & \\
%0.288201024751919`, 1.8506758344009568`, 1.0434742124414118`, 1.2261566085291755`;
%0.5661537847447308`, 1.0608301285317254`, 1.2122769429479217`, 1.9111823619996426`

\rowcolor{Apricot}%
Octopus7a & 7 / $13\pi$ & 6 & $0.963$ & $0.037$ & $0, 0.6147, 1.0574, 1.2526, 1.6722, 1.7673, 0.4758$ & $[0.553\pi, 1.447\pi]$ & 7.52 \\
\rowcolor{Apricot}%
Octopus7b & & & & & $0, 0.5093, 0.6376, 0.0647,  1.4925, 1.1079, 1.6474$ & & \\
\rowcolor{Apricot}%
Octopus7c & & & & & $0, 0.2093, 1.9577, 0.4264, 1.2674, 0.9907, 1.0733$ & & \\
%0.6146542868733116`, 1.0573841309247236`, 1.252585589048437`, 1.6722018569836692`, 1.7672558086524246`, 0.4757956457903222`;
%0.5092826327988795`, 0.6376173763976031`, 0.0646990265949169`,  1.492458792482745`, 1.107941043501673`, 1.6474499970734549`;
%0.20933638280651545`, 1.9577134971318757`, 0.42640743490400207`,  1.267380457112269`, 0.9906870638679129`, 1.0732790864209591`

Octopus9a & 9 / $17\pi$ & 8 & $0.971$ & $0.029$ & $0, 0.4497, 0.3997, 0.0660, 0.6013,$ & $[0.542\pi, 1.458\pi]$ & 9.54 \\
 & & & & & $1.3295, 1.4301, 0.9975, 1.3957$ & & \\
Octopus9b & & & & & $0, 0.2880, 0.9589, 0.8912, 1.0821,$ & & \\
 & & & & & $1.6739, 1.4258, 1.8350, 0.4227$ & & \\
Octopus9c & & & & & $0, 0.3117, 0.6148, 1.2517, 1.2243,$ & & \\
 & & & & & $0.6726, 1.2299,  1.6757, 0.0398$ & & \\
Octopus9d & & & & & $0, 0.6451, 1.0734, 1.2920, 1.6210,$ & & \\
 & & & & & $1.7451, 0.1538, 0.2252, 0.9445$ & & \\
%0.4497486648929741`, 0.3997270127832334`, 0.06596381269534715`,  0.601265485691166`, 1.329461021238165`, 1.4301305783746374`, 0.9974523499841883`, 1.3956935858441015`;
%0.2880460484346719`, 0.958914256561723`, 0.8912491450112714`, 1.082109094049384`, 1.6738707519147673`, 1.42582457223923`, 1.8350398464868998`, 0.42267538884963224`;
%0.31174163323373244`, 0.6147603006795008`, 1.2516957423622252`, 1.2242536865234077`, 0.6726441799962197`, 1.229850072395509`,  1.6756941170302067`, 0.03983234539119653`;
%0.645095902363998`, 1.0733692956230245`, 1.2920083332630472`, 1.620953575628941`, 1.7450641924221142`, 0.1538062791172177`, 0.2252176147562204`, 0.9445436425231035`
\hline
\rowcolor{Apricot}%
PB2 & 5 / 9$\pi$ & 2 & $0.922$ & $0.078$ & $0, \frac12, \frac{11}{8}, \frac{11}{8}, \frac12$ & $[0.623\pi, 1.377\pi]$ & 3.24 \\
\hline
\end{tabular}
\captionsetup{justification=raggedright,singlelinecheck=false} % align caption left
\caption{
Phases of asymmetric, altering, ultrapassband composite rotational $\theta = \pi$ sequences of $N=k/2+1$ nominal pulses ($\pi$ pulse followed by $N$ pcs of $2\pi$ pulses), which produce ultrsquare population transfer in the pulse area error sensitivity range.
The last two columns give the performance measures for UPB$N$ composite pulses --- the error correction range $[\pi (1-\epsilon_0), \pi (1+\epsilon_0)]$ wherein the errant transition probability is above the value $0.9$, and rectangularity $\varkappa$ for $\alpha = 0.1$ benchmarking. The term Octopus$N$ is attributed to UPB$N$ composite pulses because of their resemblance in appearance. 
}
\label{Table:ultrasquare-theta}
\end{table*}
%T%T%T%T%T%T%T%T%T%T%T%T%T%T%T%T%T%T%T%T%T%T%T%T%T%T%T%T%T%T%T%T%T%T%T%T%T%T%T%T%T%T%T%T%T

%T%T%T%T%T%T%T%T%T%T%T%T%T%T%T%T%T%T%T%T%T%T%T%T%T%T%T%T%T%T%T%T%T%T%T%T%T%T%T%T%T%T%T%T%T
\begin{table*}[tbph]
\begin{tabular}{|c|c|c|c|c|c|}
\hline
{\bfseries Name } & {\bfseries Pulses } & {\bfseries Number of} & {\bfseries $\sum$} & {\bfseries Phases } & {\bfseries Trace fidelity} \\
 & & {\bfseries alternations $k$} & {\bfseries {cf. \eqref{eq-z-u-bn}}} & {\bfseries $\phi_1,\phi_2, \ldots, \phi_{k/2+1}, \phi_{1} + \frac{1}{2}, \phi_{2} + \frac{1}{2}, \ldots, \phi_{k/2+1} + \frac{1}{2}$} & {\bfseries $z(\epsilon)=90\%$ } \\
 & & {\bfseries (inflection points)} &  & {\bfseries (in units $\pi$) (according to \eqref{ultrabb_phase})} & {\bfseries error correction range}\\
\hline
two & 2 & 0 & $1$ & $0$ & $[0.79517\pi, 1.20483\pi]$\\
\rowcolor{SkyBlue} %
BatPh4 & 4 & 2 & $1.(3)$ & $0, 0.6743$ & $[0.508\pi, 1.492\pi]$\\
BatPh6 & 6 & 4 & $1.5$ & $0, 0, \frac{3}{4}$ & $[0.376\pi, 1.624\pi]$\\
\rowcolor{SkyBlue}%
BatPh8 & 8 & 6 & $1.6$ & $0, 0, 0.8048, 0.6000$ & $[0.299\pi, 1.701\pi]$ \\
BatPh10 & 10 & 8 & $1.(6)$ & $0, 0, 0, 0.4129, 1.0871$ & $[0.248\pi, 1.752\pi]$\\
\rowcolor{SkyBlue}%
BatPh12 & 12 & 10 & $\approx 1.714$ & $0, 0, 0, 0.8624, 0.7142, 0.5696$ & $[0.212\pi, 1.788\pi]$\\
BatPh14 & 14 & 12 & $1.75$ & $0, 0, 0, 0, 0.8798, 0.7500, 0.6202$ & $[0.185\pi, 1.815\pi]$\\
\hline
\end{tabular}
\captionsetup{justification=raggedright,singlelinecheck=false} % align caption left
\caption{
Phases of asymmetric, altering, ultrabroadband composite phasal $\zeta = \pi$ sequences of $N=k+2$ nominal $\pi$ pulses, which produce ultrarobust Z phase gate in the ultrabroadband pulse area error correction range.
The last column gives the high-fidelity range $[\pi (1-\epsilon_0), \pi (1+\epsilon_0)]$ of pulse area error compensation wherein the trace fidelity is above the value $0.9$.}
\label{Table:ultrarobust-phase}
\end{table*}
%T%T%T%T%T%T%T%T%T%T%T%T%T%T%T%T%T%T%T%T%T%T%T%T%T%T%T%T%T%T%T%T%T%T%T%T%T%T%T%T%T%T%T%T%T

%%%%%%%%%%%%%%%%%%%%%%%%%%%%%%%%%%%%%%%%%%%%%%%%%%%%%%%%%%%%%%%%%%%%%%%%%%%%%%%%%%%%%%%%%%%%%%%%%%%%%%%%%%%%%%%%%%%%%%%%%%%%%%%%%%%%%%%%%%%%%%%%%%%%%%%%%%%%%%%%
%%%%%%%%%%%%%%%%%%%%%%%%%%%%%%%%%%%%%%%%%%%%%%%%%%%%%%%%%%%%%%%%%%%%%%%%%%%%%%%%%%%%%%%%%%%%%%%%%%%%%%%%%%%%%%%%%%%%%%%%%%%%%%%%%%%%%%%%%%%%%%%%%%%%%%%%%%%%%%%%
%%%%%%%%%%%%%%%%%%%%%%%%%%%%%%%%%%%%%%%%%%%%%%%%%%%%%%%%%%%%%%%%%%%%%%%%%%%%%%%%%%%%%%%%%%%%%%%%%%%%%%%%%%%%%%%%%%%%%%%%%%%%%%%%%%%%%%%%%%%%%%%%%%%%%%%%%%%%%%%%

%\begin{appendices}
%\end{appendices}

\end{document}